\documentclass[trackchanges,twocolumn]{aastex701}

\usepackage{siunitx}
\usepackage{amsmath}
\usepackage{tipa}

\begin{document}

\title{Flux Density Calibration of Radio Bursts Detected in the Sidelobes of the Canadian Hydrogen Intensity Mapping Experiment}

\author[0000-0003-4098-5222]{Fengqiu Adam Dong}
  \affiliation{Department of Physics and Astronomy, York University, 4700 Keele Street, Toronto, ON MJ3 1P3, Canada}
  \email{fadong@yorku.ca}
\author[0009-0008-7953-7738]{Meena Seth}
  \affiliation{Department of Physics and Astronomy, York University, 4700 Keele Street, Toronto, ON MJ3 1P3, Canada}
  \email{mmseth@yorku.ca}
\author[0000-0002-7374-7119]{Paul Scholz}
  \affiliation{Department of Physics and Astronomy, York University, 4700 Keele Street, Toronto, ON MJ3 1P3, Canada}
  \email{pscholz@yorku.ca}
\author[0000-0001-6128-3735]{Nina V.~Gusinskaia}
  \affiliation{ASTRON, Netherlands Institute for Radio Astronomy, Oude Hoogeveensedijk 4, 7991 PD Dwingeloo, The Netherlands}
  \affiliation{UvA}
  \email{gusinskaia@astron.nl}
\author[0000-0002-4279-6946]{Masui Kiyoshi W.}
  \affiliation{MIT Kavli Institute for Astrophysics and Space Research, Massachusetts Institute of Technology, 77 Massachusetts Ave, Cambridge, MA 02139, USA}
  \affiliation{Department of Physics, Massachusetts Institute of Technology, 77 Massachusetts Ave, Cambridge, MA 02139, USA}
  \email{kmasui@mit.edu}
\author[0000-0002-0772-9326]{Juan Mena-Parra}
  \affiliation{Dunlap Institute for Astronomy and Astrophysics, 50 St. George Street, University of Toronto, ON M5S 3H4, Canada}
  \affiliation{David A. Dunlap Department of Astronomy and Astrophysics, 50 St. George Street, University of Toronto, ON M5S 3H4, Canada}
  \email{juan.menaparra@utoronto.ca}
\author[0000-0002-2551-7554]{Daniele Michilli}
  \affiliation{Laboratoire d'Astrophysique de Marseille, Aix-Marseille Univ., CNRS, CNES, Marseille, France}
  \email{danielemichilli@gmail.com}
\author[0009-0003-0225-7563]{Ari Polterovich}
  \affiliation{Department of Physics, McGill University, 3600 rue University, Montr\'eal, QC H3A 2T8, Canada}
  \email{ari.polterovich@mail.mcgill.ca}
\author[0000-0001-9784-8670]{Ingrid Stairs}
  \affiliation{Department of Physics and Astronomy, University of British Columbia, 6224 Agricultural Road, Vancouver, BC V6T 1Z1 Canada}
  \email{stairs@astro.ubc.ca}
\newcommand{\allacks}{
We acknowledge that CHIME is located on the traditional, ancestral, and unceded territory of the Syilx/Okanagan people. We are grateful to the staff of the Dominion Radio Astrophysical Observatory, which is operated by the National Research Council of Canada. CHIME operations are funded by a grant from the NSERC Alliance Program and by support from McGill University, University of British Columbia, and University of Toronto. CHIME was funded by a grant from the Canada Foundation for Innovation (CFI) 2012 Leading Edge Fund (Project 31170) and by contributions from the provinces of British Columbia, Québec and Ontario. The CHIME/FRB Project was funded by a grant from the CFI 2015 Innovation Fund (Project 33213) and by contributions from the provinces of British Columbia and Québec, and by the Dunlap Institute for Astronomy and Astrophysics at the University of Toronto. Additional support was provided by the Canadian Institute for Advanced Research (CIFAR), the Trottier Space Institute at McGill University, and the University of British Columbia. The CHIME/FRB baseband recording system is funded in part by a CFI John R. Evans Leaders Fund award to IHS.

We acknowledge that CHIME and the \textbf{k’ni\textipa{P}atn k’l$_\smile$stk’masqt} Outrigger (KKO) are built on the traditional, ancestral, and unceded territory of the Syilx Okanagan people. \textbf{k’ni\textipa{P}atn k’l$_\smile$stk’masqt} is situated on land leased from the Imperial Metals Corporation. We are grateful to the staff of the Dominion Radio Astrophysical Observatory, which is operated by the National Research Council of Canada. CHIME operations are funded by a grant from the NSERC Alliance Program and by support from McGill University, University of British Columbia, and University of Toronto. CHIME/FRB Outriggers are funded by a grant from the Gordon \& Betty Moore Foundation. We are grateful to Robert Kirshner for early support and encouragement of the CHIME/FRB Outriggers Project, and to Dusan Pejakovic of the Moore Foundation for continued support. CHIME was funded by a grant from the Canada Foundation for Innovation (CFI) 2012 Leading Edge Fund (Project 31170) and by contributions from the provinces of British Columbia, Québec and Ontario. The CHIME/FRB Project was funded by a grant from the CFI 2015 Innovation Fund (Project 33213) and by contributions from the provinces of British Columbia and Québec, and by the Dunlap Institute for Astronomy and Astrophysics at the University of Toronto. Additional support was provided by the Canadian Institute for Advanced Research (CIFAR), the Trottier Space Institute at McGill University, and the University of British Columbia. The CHIME/FRB baseband recording system is funded in part by a CFI John R. Evans Leaders Fund award to IHS.

F.A.D is a Canadian SKA Scientist and is funded by the Government of Canada / est financé par le gouvernement du Canada.
P.S. acknowledges the support of an NSERC Discovery Grant (RGPIN-2024-06266).
K.W.M. is supported by NSF Grant No. 2510771.
J.M.P. acknowledges the support of an NSERC Discovery Grant (RGPIN-2023-05373).
D.M. acknowledges support from the French government under the France 2030 investment plan, as part of the Initiative d'Excellence d'Aix-Marseille Universit\'e -- A*MIDEX (AMX-23-CEI-088).
FRB research at UBC is supported by an NSERC Discovery Grant and by the Canadian Institute for Advanced Research.  
}

\begin{abstract}
The Canadian Hydrogen Intensity Mapping Experiment/Fast Radio Burst (CHIME/FRB) instrument is a transient astrophysical burst survey that is discovering FRBs, pulsars, and long-period transients. As a transit telescope, the survey is most sensitive within 2$^\circ$ of the meridian; however, CHIME/FRB can detect bright astrophysical bursts out to its horizon, at $10^{-2}$--$10^{-3}$ of its peak sensitivity. In this study, we detail the flux density calibration procedure for these bursts detected in the CHIME sidelobes for both channelized voltage (baseband) and intensity data. We validate and estimate the calibration uncertainties by beamforming baseband data towards bright, steady sources in the sidelobes, comparing baseband and intensity flux density calibrations of the same astrophysical transients, and calibrating intensity flux densities for transiting steady sources. We are able to calibrate flux densities for declinations spanning $-12^\circ$ to $+59^\circ$ out to hour angles of 30$^\circ$, and out to 90$^\circ$ in hour angle for declinations near that of Taurus A (Crab Nebula, $\delta\approx+22^\circ$). We conclude that uncertainties in the flux density calibration for transients in the CHIME sidelobes range from 20\% to 42\%, depending on the data used and the hour angle of the burst.
\end{abstract}



\section{Introduction}
Fast radio bursts (FRBs) are bright millisecond-duration radio bursts that occur at extragalactic distances. The brightest reach flux densities in excess of 1000\,Jy\citep{10.3847/2041-8213/adf62f}. Moreover, Galactic counterparts to FRBs are generally extremely bright. For example, SGR\,1935+2154 emitted bursts with peak flux densities between 110--150\,kJy at 600\,MHz \citep{10.1038/s41586-020-2863-y,Bochenek:Ravi:Belov:2020}.

The Canadian Hydrogen Intensity Mapping Experiment (CHIME) is a commensal transit instrument hosting the CHIME/Fast Radio Burst (CHIME/FRB) project. While it has a large survey volume in the main lobe, covering $\sim$200 square degrees \citep{10.3847/1538-4357/aad188}, the field of view can be much larger if one conducts a survey at lower sensitivity. For example, at an hour angle offset of $\sim25^{\circ}$ from the meridian, CHIME's sensitivity is $\sim 10^{-3}$ of its peak value \citep{lin:scholz:ng:2024}. Therefore, the CHIME/FRB survey, when including the sidelobes, can also be considered a low-sensitivity survey with horizon-to-horizon coverage. This enables detection of extremely bright, albeit rare, radio bursts, such as those from nearby FRBs or bright Galactic FRB counterparts. Several new instruments aim for comparable all-sky coverage by design, using wide-field antenna arrays, such as the Bustling Universe Radio Survey Telescope in Taiwan (BURSTT; \citealt{lin:lin:li:2022}) or the Coherent All-Sky Monitor (CASM; \citealt{connor:ravi:sanghavi:2026}).

In this study, we describe the flux density calibration procedure for bursts detected in the sidelobe regime of CHIME/FRB. In Section \ref{sec:methods}, we describe the method for the flux density calibration of both channelized raw voltage (baseband) and total-intensity data. In Section \ref{sec:validation}, we present validation checks on each variable in the flux density calibration procedure and provide uncertainties. Finally, in Section \ref{sec:conclusions}, we discuss the conclusions and future work. 

\section{Methods}
\label{sec:methods}

\begin{figure*}[tbp]
  \centering  
  \includegraphics[width=0.48\textwidth]{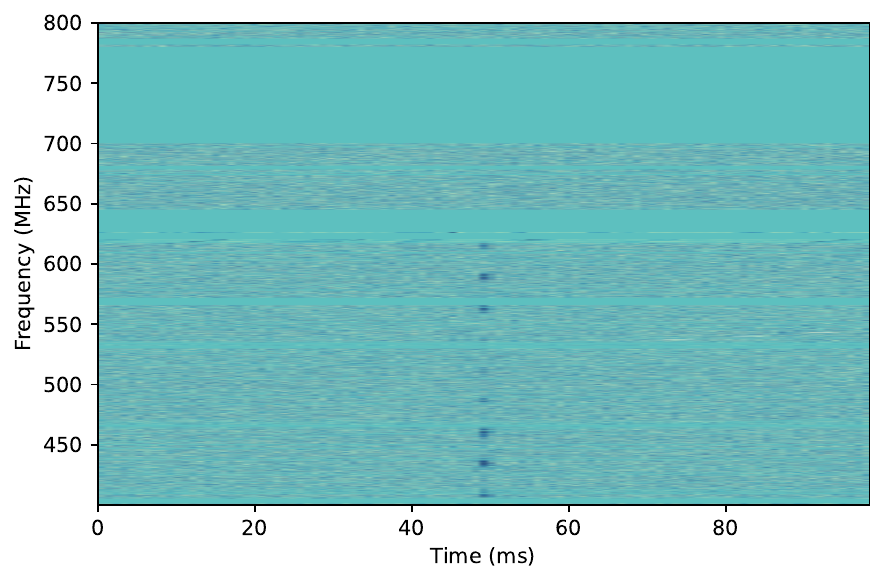}
  \includegraphics[width=0.48\textwidth]{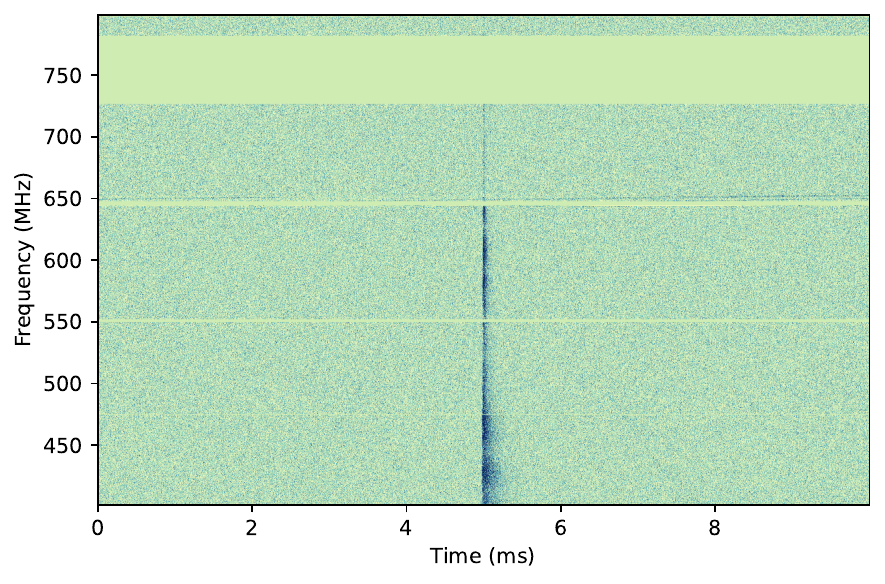}
  \caption{The dynamic spectrum of a Crab giant pulse detected by CHIME/FRB in the far sidelobe on 2025 July 05, shown for beamformed intensity (left) and baseband (right) data. Channels with constant values have been masked for RFI. Many fewer channels were masked for the baseband data than for the intensity data, as the baseband data are beamformed directly towards the Crab pulsar while the intensity data are beamformed in the primary beam of CHIME. We subtracted the median for each channel and divided by the standard deviation for both figures. The intensity data are ``notched'' by the FFT-formed beam response in the far sidelobes; this structure is not intrinsic to the pulse, and it is corrected in the baseband data. Much of this structure is hidden by the RFI masking in the intensity panel. Note that the two panels have different time spans: the intensity data cover 90\,ms and the baseband data cover 9\,ms.}
  \label{fig:baseband_vs_intensity}
\end{figure*}

Within the CHIME primary beam field of view, the CHIME/FRB instrument forms 1024 regularly spaced Fast Fourier Transform (FFT) beams \citep{10.23919/URSIGASS.2017.8105318,Masui:Shaw:Ng:2019}. These beams are arranged in 256 North-South rows with 4 East-West columns. The CHIME/FRB instrument then searches each individual beam for astrophysical transients. Throughout this work, ``sidelobe'' refers to the sidelobes of the primary beam, i.e., sky positions more than $\sim2^{\circ}$ from the meridian. Although the FFT-formed beams tile the primary beam field of view, bursts at these positions are detected at reduced sensitivity through the sidelobe response of the formed beams. These transients include FRBs \citep{10.3847/1538-4365/ac33ab,10.3847/1538-4365/ae3828}, pulsars \citep{Good:Andersen:Chawla:2021,Dong:Crowter:Meyers:2023}, and, more recently, long-period transients \citep{Dong:Clarke:Curtin:2025,Dong:Shin:Law:2025}. For each radio transient detected, two data products are stored. The first type of data consists of 0.983\,ms Stokes I (intensity) data from each FFT-formed beam that recorded a signal with S/N$>$8.5. The intensity data include 16384 evenly spaced frequency channels between 400--800\,MHz. The second type of data product, channelized voltage (baseband), is saved only when S/N$>$12. Baseband data contain polarization and phase information of radio waves measured by each antenna of the array at the native resolution of about 2.56\,$\mu$s. Baseband data are used to form a digital beam to the source direction, which maximizes the telescope's sensitivity and removes the FFT-formed beam response, leaving only the effect of the primary beam. We provide an example of both types of data products in Figure \ref{fig:baseband_vs_intensity}.
 
In this section, we present the calibration procedures for bursts detected in the CHIME/FRB sidelobes, in both intensity and baseband data. The validation and error analysis of these methods are provided in the subsequent sections.

\subsection{Primary and Formed Beam Models}
\begin{table*}[tbp]
\caption{Table of holography sources and the dates on which holography data were taken.}
\centering
\begin{tabular}{lccc}
\hline\hline
Source      & Declination & Max Hour Angle ($^\circ$) & Observation Dates \\
\hline\hline\\
Cassiopeia A & +58:48:41 & $\pm30$ & 2018 Aug 10, 11, 14--16\\
Cygnus A    & +40:44:02  & $\pm30$ & 2018 Sep 6, 7, 9--13, 15, 16, 19\\
Taurus A    & +21:59:05  & $\pm90$ & 2018 Oct 14, 24\\
Virgo A     & +12:23:28  & $\pm30$ & 2018 Sep 11, 12, 15, 18, 21--23; 2020 Jan 10\\
Hercules A  & +04:59:34  & $\pm30$ & 2018 Nov 4, 18--20, 24, 25\\
Hydra A     & -12:05:43  & $\pm30$ & 2019 Jan 3, 5, 6, 8, 9, 11--13, 15, 16\\
\hline\hline
\end{tabular}
\label{tab:holo_sources}
\end{table*}
\begin{figure*}[tbp]
  \centering
  \includegraphics[width=0.45\textwidth]{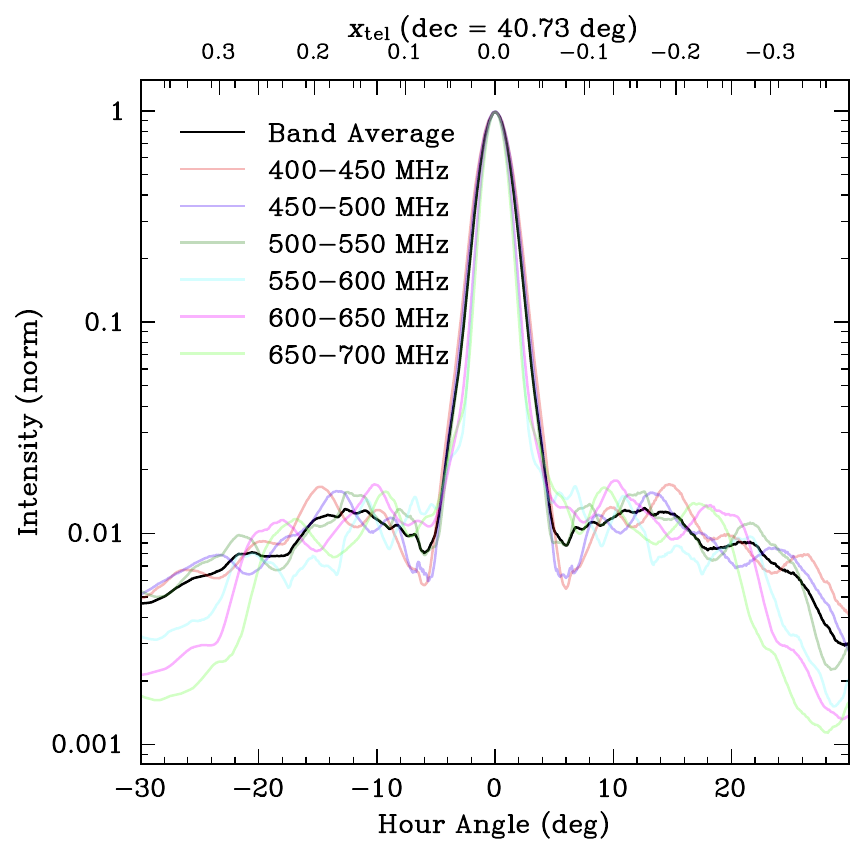}
  \includegraphics[width=0.45\textwidth]{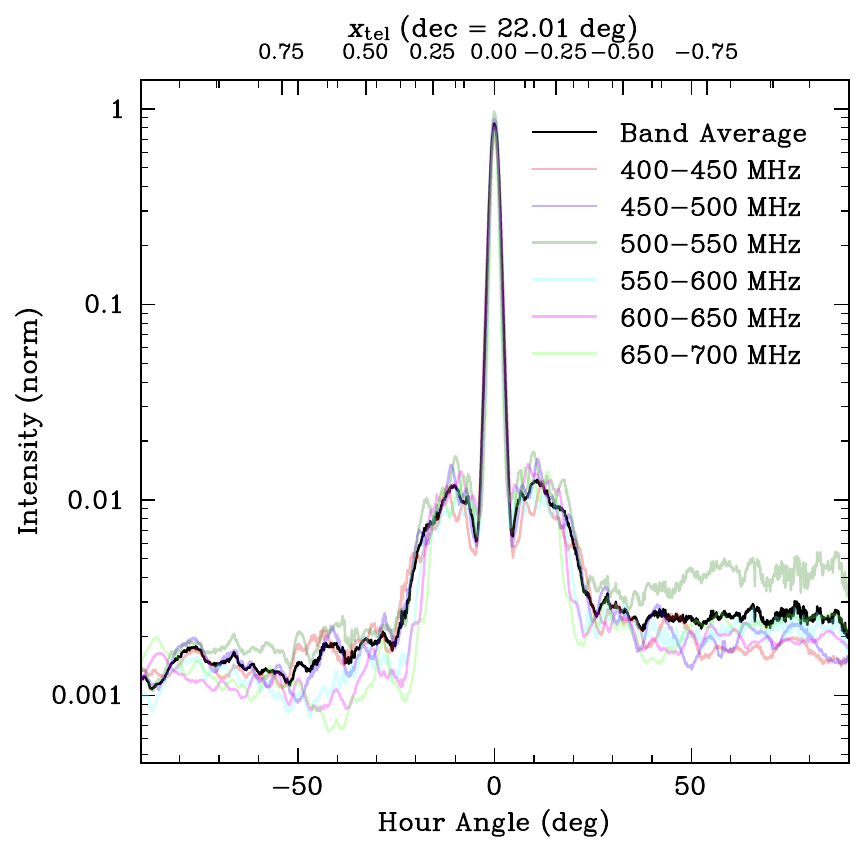}
  \caption{The holography beam model for Cygnus A (left) and Taurus A (right). The beam models are normalized to Cygnus A at transit, as is conventional for CHIME beam models. Here, we provide the band-averaged response and also the response in 50\,MHz bins. We omit 700--800\,MHz because these frequencies are largely contaminated by RFI; some residual RFI may also affect the highest-frequency bin at large hour angles.}
  \label{fig:holography_primary}
\end{figure*}
Throughout this study, we employ the primary and FFT-formed beam models to characterize CHIME/FRB's response to an astrophysical source. The primary beam model describes the response of each individual feed as a function of sky position. By convention, the standard primary beam model is normalized to the position of Cygnus A at transit such that all frequencies between 400--800\,MHz have a primary beam model response of 1. The FFT-formed beam model describes the response of each of the 1024 FFT-formed beams generated within the CHIME/FRB main lobe as a function of sky position. The FFT-formed beam models are normalized by the peak of the beam-center response over frequency; that is, the formed beam response equals 1 at the beam center for only some frequencies.

The standard primary-beam model used for flux density calibration by CHIME/FRB \citep{Merryfield:Tendulkar:Shin:2023,Andersen:Patel:Brar:2023,10.3847/1538-4357/ad464b} is only reliable within about 2$^\circ$ of the meridian \citep{10.3847/1538-4357/ad8133,10.3847/1538-4365/ac6fd9}. Therefore, we employ holography techniques to characterize CHIME's primary beam response in the sidelobes. Specifically, bright radio sources are observed with the 26\,m John A. Galt Telescope at the Dominion Radio Astrophysical Observatory (hereafter the Galt Telescope) for at least 4 hours around the CHIME transit. The signal is cross-correlated with each of the 1024 individual CHIME feeds. Because the Galt Telescope tracks the source, its own response to that source is constant, so the cross-correlation directly measures CHIME's primary beam. Further details for generating the holography dataset are provided in \cite{10.3847/1538-4357/ad8133}. This technique was previously used in \cite{lin:scholz:ng:2024} to compute an equivalent on-axis S/N for FRBs discovered in sidelobes. Here, we take this one step further and use the holography beam model to estimate the flux density of the detected bursts. To generate a reliable primary beam response across all hour angles, we average the correlation response over all individual feeds, as produced in \cite{10.3847/1538-4357/ad8133}. Cylinder B is excluded from this process because previous studies have found an overall systematic shift in holography correlation strength due to slight misalignment of the feeds \citep{10.3847/1538-4357/ad8133}. The resulting beam response is given in Figure \ref{fig:holography_primary} for Cygnus A and Taurus A. In Table \ref{tab:holo_sources} we list all the sources for which we have produced a holography beam model. We emphasize that the holography primary beam used throughout this work is a distinct, independently measured model; the validation presented in Section \ref{sec:validation} therefore tests the holography primary beam rather than the standard, normalized primary-beam model.
\begin{figure*}[tbp]
  \centering
  \includegraphics[width=0.48\textwidth]{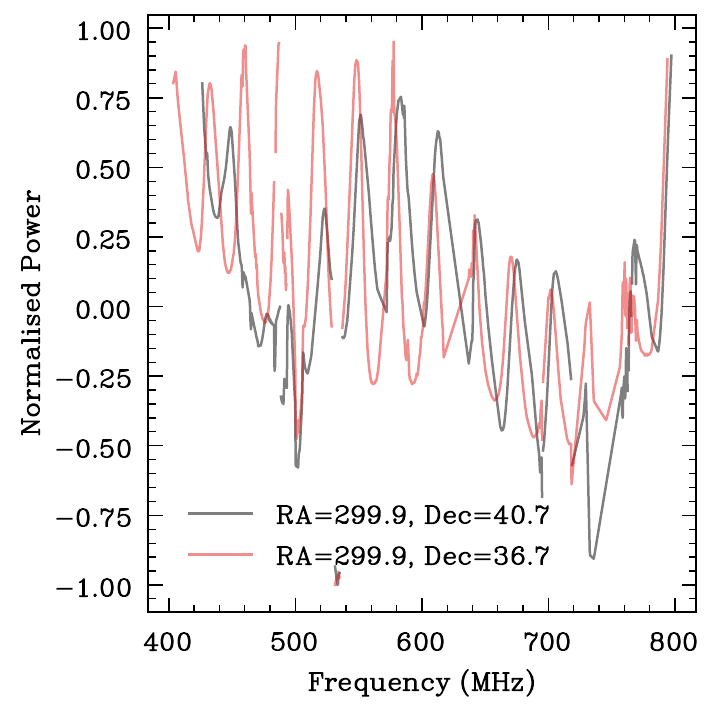}
  \includegraphics[width=0.48\textwidth]{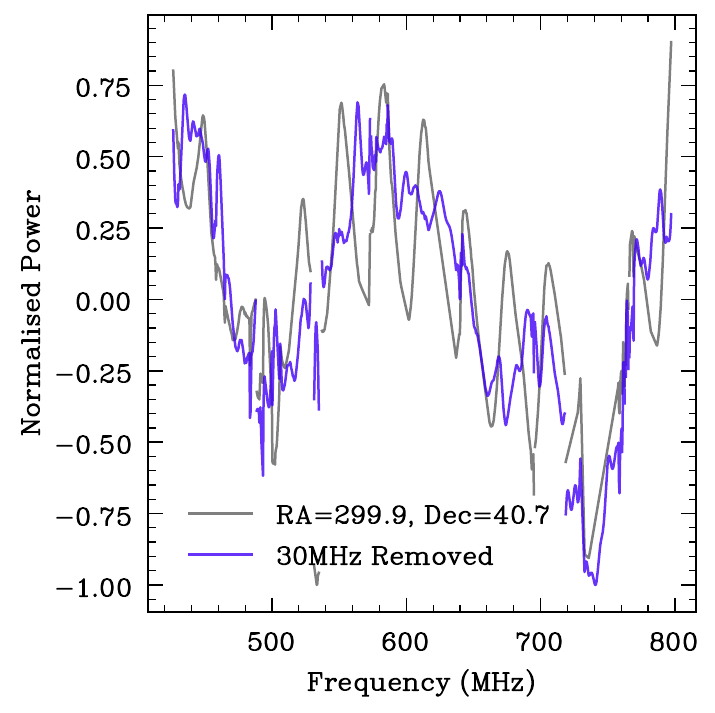}
  \caption{Left: the CHIME bandpass measured from baseband data beamformed to the position of Cygnus A while it was in the sidelobe. Overlaid is the bandpass beamformed to a position offset by $4^{\circ}$ in declination from Cygnus A. The 30\,MHz ripples of the two spectra are not aligned because of the declination offset. Each spectrum is normalized by its maximum absolute value before RFI masking, so that all values lie between $-1$ and 1. Right: the same Cygnus A bandpass, with the dotted line showing it after the 30\,MHz ripple and all of its harmonics have been masked in the Fourier domain. While low-amplitude fluctuations remain, they are no longer a cause of division-by-small-number instability.}
  \label{fig:30MHz}
\end{figure*}

The CHIME primary beam response exhibits a $\sim30$\,MHz sinusoidal ripple across frequency \citep{10.3847/1538-4365/ac6fd9}. This is due to the instrument's optical design. Incoming light paths can reflect along the cylinder's long axis, creating resonances at specific frequencies. The sinusoid also depends on the sky location of interest. This effect is well modeled in the main lobe and can thus be readily corrected during flux density calibration; however, it is poorly understood in the sidelobes. Because only a handful of sources are bright enough to yield holography data sets, the primary beam is sparsely sampled in declination, so any offset between the transient declination and the holography track's declination causes the 30\,MHz sinusoid to be misaligned. This, in turn, leaves large residual ripples in the calibrated spectrum, because the calibration divides by near-zero beam values. This effect is demonstrated in Figure \ref{fig:30MHz}. We remove the sinusoid from the primary beam by attenuating its response in the Fourier domain. For each hour angle slice presented in Figure \ref{fig:holography_primary}, we take the Fourier transform of the frequency response and mask the peaks induced by the $30$\,MHz sinusoid and its first five harmonics. We are therefore left with a slowly varying primary beam response across the frequency range. This can be seen in the right panel of Figure \ref{fig:30MHz}.

Finally, we transform the hour angle and declination of the source into telescope coordinates. This step is required because sources at different declinations trace different great circles across the CHIME sky, so the same hour angle offset from the meridian corresponds to a different angular distance from it, and therefore to a different beam attenuation. Taken to the extreme, a source at a declination near 90$^\circ$ remains close to the CHIME meridian at all hour angles and is therefore always observed at relatively high sensitivity, whereas a source at low declination moves away from the meridian quickly and its sensitivity changes rapidly with hour angle. We adopt the same transformation as \cite{10.3847/1538-4365/ac6fd9}, projecting the coordinates onto a unit sphere, using their Equations 1 and 2:
\begin{equation}
  x_{\mathrm{tel}} = -\cos\delta\,\sin h
  \label{eq:xtel}
\end{equation}
\begin{equation}
  y_{\mathrm{tel}} = \cos\delta_{0}\,\sin\delta - \sin\delta_{0}\,\cos\delta\,\cos h,
  \label{eq:ytel}
\end{equation}
where $h$ is the hour angle of the source, $\delta$ is its declination, and $\delta_{0}=+49.3^\circ$ is the latitude of CHIME. $x_{\mathrm{tel}}$ and $y_{\mathrm{tel}}$ are the orthographic projections, centered on the zenith, of the unit vector pointing towards the source; $x_{\mathrm{tel}}$ is parallel to the East--West direction and $y_{\mathrm{tel}}$ to the North--South direction. Working in $x_{\mathrm{tel}}$ allows a holography track measured at one declination to be applied to a source at another. Note that while the underlying implementation adopts the telescope coordinate transformation, here we present everything in terms of hour angle for easier comprehension and readability.

The FFT-formed beams are created based on the position of the feeds; therefore, we can model the formed beam responses at any point above the horizon for any of the 1024 FFT-formed beams analytically \citep{Masui:Shaw:Ng:2019}. The response is highly chromatic, and the beam centers drift north-south as a function of observing frequency. By modeling the FFT-formed beam response and combining it with the primary beam, we can produce a ``total'' CHIME beam response. This combined response includes the chromatic effects of both the primary and formed-beam models.

\subsection{Intensity Calibration}
\begin{figure*}[tbp]
  \centering  \includegraphics[width=0.85\textwidth]{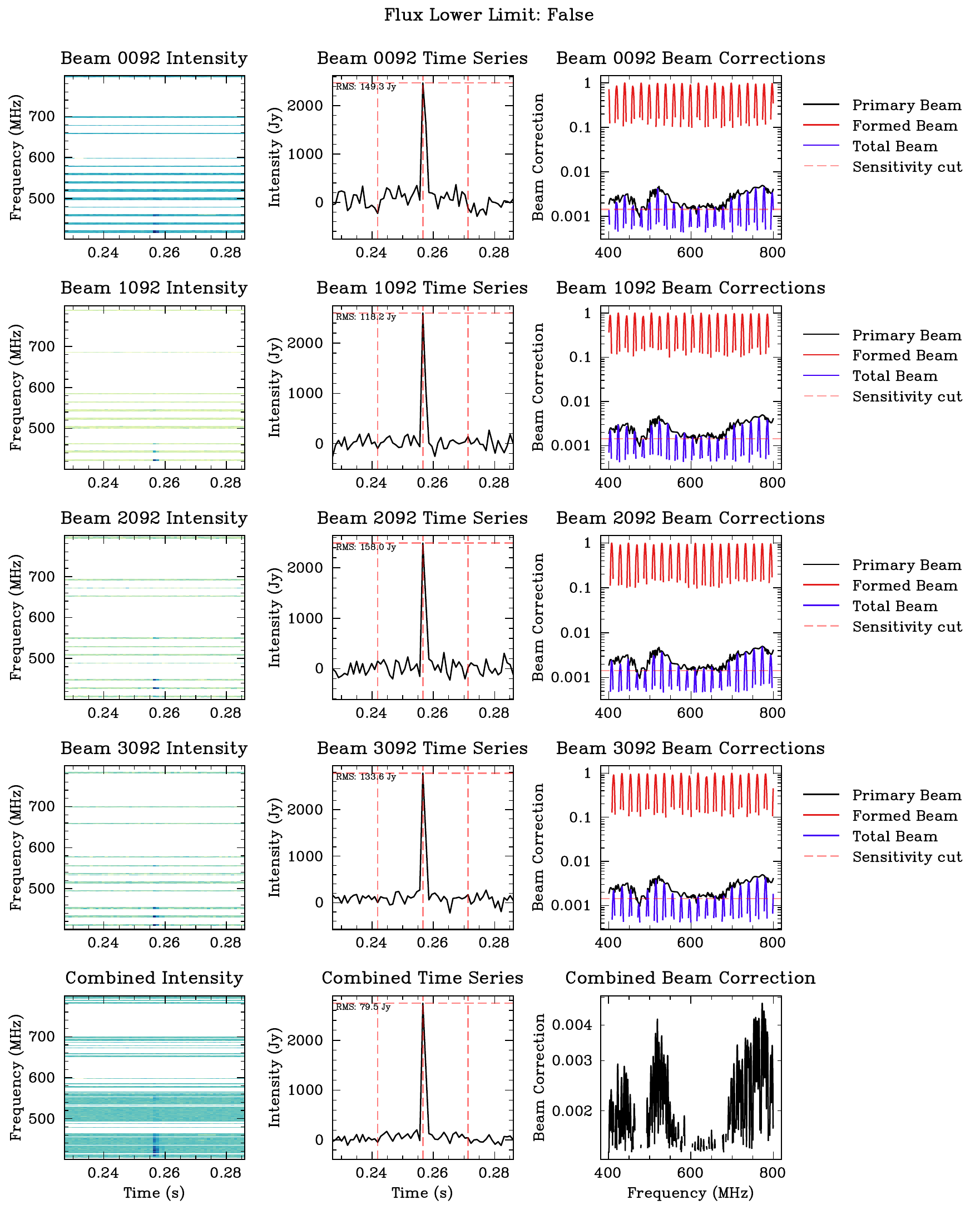}
  \caption{An example diagnostic plot of a calibrated burst emitted by PSR B0531+21. Intensity data at 0.983\,ms time resolution were used to generate this figure. The left panels show the dynamic spectrum of the burst with known RFI and low-sensitivity channels masked. The middle panels show the dedispersed and frequency-averaged time series. The right panels show the modeled beam response, i.e., the correction applied in Equation \ref{eq:intensity_fluxcal}. Each row corresponds to one detection beam, given in the panel titles. The final row shows the frequency-spliced combination of all the beams, which improves the burst S/N. The overall title states whether the flux density calibration is a lower limit.}
  \label{fig:example_fluxcal}
\end{figure*}
In the past, calibrating CHIME data relied on steady sources with known flux densities, which are frequently monitored \citep{Andersen:Patel:Brar:2023}. However, \citet{Andersen:Patel:Brar:2023,Merryfield:Tendulkar:Shin:2023} showed that CHIME/FRB intensity data are already flux-calibrated due to daily phase and amplitude calibration during beamforming. Therefore, the intensity values recorded by CHIME/FRB are related to the flux density by a constant factor and the primary and FFT-formed beam model responses. The flux density calibration equation is
\begin{equation}
  \begin{split}
    S_{\mathrm{src}}(\nu,\alpha,\delta,t) =& \frac{1024 f_{\mathrm{good}}^{2} \cdot 128}{4^{2}\cdot 0.806745 \cdot 100}\\
                                  &\times\frac{1}{M_{\mathrm{primary}}(\nu,\alpha,\delta,t) \cdot M_{\mathrm{FFT}}(\nu,\alpha,\delta,t)}\\
                                  &\times I_{\mathrm{BF}}
    \end{split}
    \label{eq:intensity_fluxcal}
\end{equation}
where $f_{\mathrm{good}}$ is the fraction of good feed inputs, $M_{\mathrm{primary}}$ is the holography beam model, $M_{\mathrm{FFT}}$ is the FFT-formed beam model, $I_{\mathrm{BF}}$ is the intensity in beamformer units, and $S_{\mathrm{src}}$ is the flux density in Jy of a source. The number of good feed inputs varies daily, often due to weather conditions, and ranges from 70--95\%\footnote{After 2020-Apr-23, this correction is no longer needed; instead, it has been applied at the gain calibration step.}. The constant factors arise from corrections and scaling applied to the data during beamforming and upchannelization. These numerical factors derive from the internal beamformer and upchannelizer scaling and are not publicly documented.

Bursts detected in the far sidelobes exhibit a pronounced ``lobed/notched'' structure. This is due to the frequency response of the formed beams, as shown in Figure \ref{fig:baseband_vs_intensity}. If we calibrated every frequency channel and then averaged over the band, the channels with low beam response would dominate the result, because dividing by a small $M_{\mathrm{FFT}}$ in Equation \ref{eq:intensity_fluxcal} amplifies their noise along with their signal. Consequently, in the flux density calibration routine, we mask all channels with a response less than 10\% of the peak response. This is shown in Figure \ref{fig:example_fluxcal} by the missing channels in the dynamic spectrum and the right-most beam response panels.

\subsection{Frequency Splicing}
When only intensity data are recorded for sidelobe bursts, we take advantage of the fact that multiple beams are usually triggered to boost S/N of the resulting burst. As seen in Figure \ref{fig:example_fluxcal}, the formed beam response consists of many spikes. The frequency at which the response peaks differs from beam to beam; therefore, we can combine all the beams into one dynamic spectrum. For each frequency channel, we pick the beam with the most sensitivity and produce an aggregate dynamic spectrum. This is shown in the last panel of Figure \ref{fig:example_fluxcal}, where the baseline noise improves by a factor of $\sim$2 (an average RMS of 140\,Jy per beam versus 70\,Jy for the combined beam). This splicing procedure increases the S/N of the resulting flux calibration; while it has little effect on the systematic errors described in later sections, it reduces the statistical errors (i.e., the off-pulse noise) significantly.

\subsection{Baseband Calibration}
\label{sec:baseband}
The beamformed baseband data are also related to the flux density by a constant factor and the primary beam model. This assumes that the position of the source is well known, and that the phased-array beam used in the baseband data analysis process \citep{Michilli:Masui:Mckinven:2021} is formed at the position of the astrophysical source as localized by the baseband pipeline \citep{10.3847/1538-4357/ad464b}. 

There are many benefits to baseband data over intensity data, at the cost of much larger storage requirements. One major advantage of baseband data for bursts detected in the sidelobes is that they can recover the signal across the 400--800\,MHz band that was lost due to the response of the FFT-formed beam, significantly increasing S/N, as shown in Figure \ref{fig:baseband_vs_intensity}.

The relationship between baseband beamformed intensity and flux density is given by
\begin{equation}
  S_{\mathrm{src}}(\nu)=\frac{2\left[I_{\mathrm{on}}(\nu)-I_{\mathrm{off}}(\nu)\right]}{F_{\mathrm{good}}^2 \cdot M_{\mathrm{primary}}(\nu)}
  \label{eq:baseband_fluxcal}
\end{equation}
where $S_{\mathrm{src}}(\nu)$ is the source flux density at frequency $\nu$, $I_{\mathrm{on}}(\nu)$ and $I_{\mathrm{off}}(\nu)$ are the on- and off-source beamformed intensities, and $M_{\mathrm{primary}}(\nu)$ is the frequency-dependent holography primary beam model. $F_{\mathrm{good}}$ is the number of good feed inputs contributing to the beamforming process, a number between 0 and 2048 (1024 feeds $\times$ 2 polarizations); it is related to the fraction used in Equation \ref{eq:intensity_fluxcal} by $F_{\mathrm{good}}=2048f_{\mathrm{good}}$. For bursts, we use an off-pulse region in time for $I_{\mathrm{off}}(\nu)$; for steady sources, we beamform away from the source location.\footnote{After 2020-Apr-23, $I_{\mathrm{on,\,off}}$ is implicitly scaled when beamforming, such that the output of the beamforming process is scaled as if all feeds contribute to it. Therefore, after 2020-Apr-23, $F_{\mathrm{good}}=2048$.}
  
\section{Validation and Uncertainties}
\label{sec:validation}
To test the flux density calibration, we split the validation procedures into three separate sections. First, we evaluate the holography primary beam response. This tests the largest source of uncertainty in Equations \ref{eq:intensity_fluxcal} and \ref{eq:baseband_fluxcal}, $M_{\mathrm{primary}}$. We do this by beamforming baseband data on and off a bright, steady source across a range of hour angles. Second, we evaluate the FFT beam model, $M_{\mathrm{FFT}}$, which is the second-largest source of uncertainty. We do this by flux density calibrating the same set of astrophysical pulses in the far sidelobes with both intensity and baseband data. Because the same holography primary beam model $M_{\mathrm{primary}}$ multiplies both the baseband and intensity calibrations, it cancels in the fluence ratio. The ratio therefore isolates the FFT-formed beam model $M_{\mathrm{FFT}}$, and the systematic uncertainty on $M_{\mathrm{primary}}$ does not propagate into this validation. Finally, we evaluate the constant factors in Equation \ref{eq:intensity_fluxcal}. We do this by flux density calibrating steady sources when they are in the main lobe of CHIME.

\subsection{Validation of the Holography Beam Model}
\label{sec:holo}
\begin{figure}[tbp]
  \centering
  \includegraphics[width=0.45\textwidth]{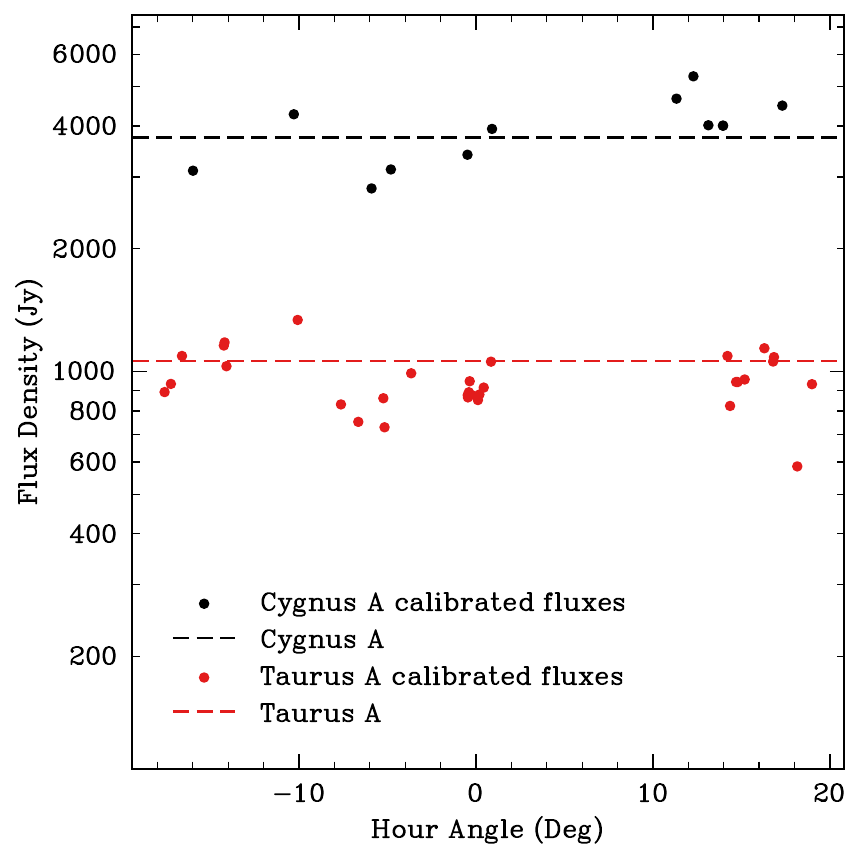}
  \caption{The band-averaged flux density values of Cygnus A and Taurus A. This was derived by beamforming the baseband data on and off-source. The dashed lines show the respective catalog flux densities in the CHIME frequency band. There is a gap between $\sim$0--13$^\circ$ in Cygnus A because of a secondary bright source that transits at that time, biasing flux calibration results. Gaps in the Taurus A hour angle range are due to a lack of usable baseband data. }
  \label{fig:holography_validation}
\end{figure}
To evaluate the holography primary beam model, we first query the CHIME/FRB database for all archival baseband data recorded while a target steady source was within 20$^\circ$ in hour angle of the CHIME meridian. These data were saved by transient triggers during the CHIME/FRB survey. Each baseband data segment has a duration of 100\,ms with a time resolution of 2.56\,$\mu$s. For this study, we selected Cygnus A and Taurus A to validate the holography beam model. We beamform on and off the target source, using a 4$^\circ$ declination offset for the off-source pointing; this offset is well beyond the primary beam's full width at half maximum of $\sim0.5^\circ$, so the off-source beam contains negligible source signal. We then follow Equation \ref{eq:baseband_fluxcal} to produce a band-averaged flux density. The results are shown in Figure \ref{fig:holography_validation}. The calibrated Cygnus A data have a median flux density of $S_{\mathrm{CygA}}=3677$\,Jy and a root-mean-square error ($\mathrm{RMSE}$) of 603\,Jy, relative to the catalog flux density \citep{perley:butler:2017}. These correspond to a median ratio of 0.98 and a fractional $\mathrm{RMSE}$ of 0.16. For Taurus A, we find a median flux density of $S_{\mathrm{TauA}}=932$\,Jy and an $\mathrm{RMSE}$ of 155\,Jy, corresponding to a median ratio of 0.93 and a fractional $\mathrm{RMSE}$ of 0.16 \citep{perley:butler:2017}. Here the median ratio is $\mathrm{median}(S_{\mathrm{calibrated}}/S_{\mathrm{catalog}})$, the $\mathrm{RMSE}$ is the root-mean-square difference between the calibrated and catalog flux densities, and the fractional $\mathrm{RMSE}$ is that difference divided by the catalog flux density.

Two effects that we cannot test directly may inflate these uncertainties for a real transient. First, a detected sidelobe burst will generally not lie exactly at the declination of a holography source. Second, the burst may occur at an hour angle greater than 20$^\circ$; beyond this, most steady sources are too faint for their flux density to be retrieved reliably from a single 100\,ms baseband segment, so the beam model is untested there. We therefore round the measured fractional scatter of 0.16 up to 0.2 for sources within 20$^\circ$ in hour angle, and conservatively double it to 0.4 beyond 20$^\circ$, as the fractional uncertainty due to the holography primary-beam model.
\subsection{Validation of the FFT-Formed Beam Model in the Far Sidelobes}
\label{sec:MFFT}
\begin{figure}[tbp]
  \centering
  \includegraphics[width=0.45\textwidth]{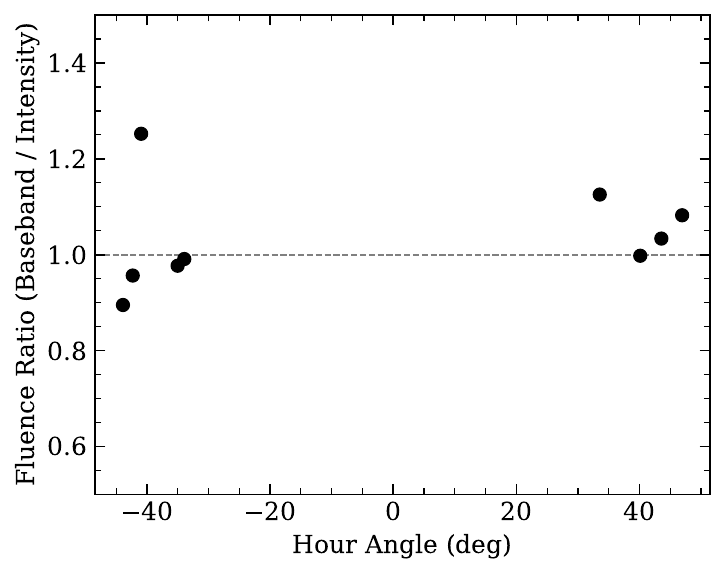}
  \caption{The ratio of the fluence of Crab pulsar pulses calibrated using baseband data to that calibrated using intensity data, as a function of hour angle.}
  \label{fig:fluence_ratio}
\end{figure}
In this section, we test the validity of the formed beam model $M_{\mathrm{FFT}}$ in the far sidelobes. This is crucial for understanding the flux density calibration of far sidelobe events when baseband data are unavailable. To evaluate the formed beam model, we identify nine pulses from the Crab pulsar that were detected in the far sidelobes and for which both intensity and baseband data are available. The data for each pulse are then flux-calibrated for both the baseband and intensity data using the methods described in Section \ref{sec:methods}. Because the intensity spectrum is notched, we keep only those frequency channels that survive the intensity-data masking when calibrating the baseband data, so that both data sets cover the same channels. We then compute the band-averaged flux density time series for both the intensity and baseband data. Finally, we integrate over the pulse duration to find the fluence. The fluence ratios are shown in Figure \ref{fig:fluence_ratio}. The median ratio, median$(F_{\mathrm{baseband}}/F_{\mathrm{intensity}})$, is 0.998, with a fractional $\mathrm{RMSE}$ of 0.106. The baseband and intensity calibrations therefore agree to within 0.2\%, with the fractional RMSE of 0.106 taken as the uncertainty on $M_{\mathrm{FFT}}$.

\subsection{Validation of Intensity Calibration Constants}
\label{sec:intensity}
\begin{figure}[tbp]
  \centering
  \includegraphics[width=0.45\textwidth]{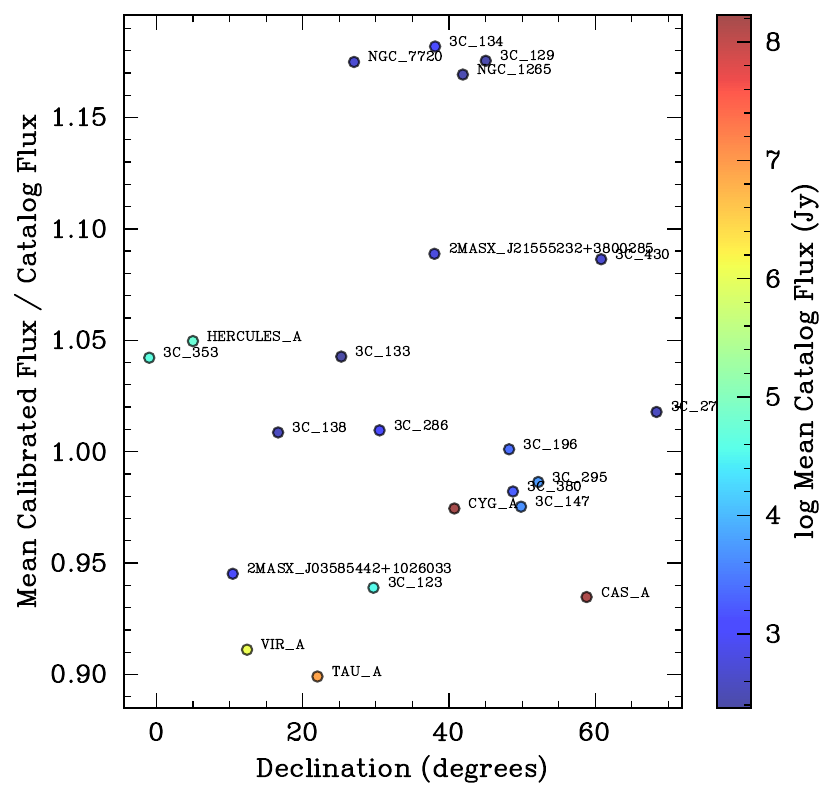}
  \caption{Intensity flux density calibration of 22 steady sources as they transit CHIME. The data are recorded at the CHIME meridian. We show the ratio of our band-averaged flux density to the catalog value.}
  \label{fig:intensity_validation}
\end{figure}
Finally, we test the constant factors in Equation \ref{eq:intensity_fluxcal}. To do this, we take 22 steady sources that transit the CHIME telescope and calculate their flux density in the main lobe as they traverse the FFT-formed beam that corresponds to CHIME's meridian. These data were collected during the original steady-source calibration process outlined in \cite{Andersen:Patel:Brar:2023}. The data consist of 0.983\,ms intensity data from a single FFT-formed beam, collected near-daily between 23 September 2018 and 31 December 2025. Each steady source is then flux density calibrated using Equation \ref{eq:intensity_fluxcal}. The results are provided in Figure \ref{fig:intensity_validation}. The median ratio between calibrated flux densities and the catalog flux densities, $\mathrm{median}(S_\mathrm{calibrated}/S_\mathrm{catalog})$, is 1.01, with a fractional $\mathrm{RMSE}$ of 0.07.

\subsection{Combining uncertainties}
The uncertainties from each variable in this section are treated as independent; therefore, the fractional uncertainties can be added in quadrature. We summarize the uncertainties in Table \ref{tab:total_error}. The dominant source of uncertainty is the holography beam model. This is followed by the FFT beam model for intensity data, and the flux calibration constants contribute the least to the total uncertainty budget. Furthermore, note that when performing flux density calibrations, we also add baseline Gaussian noise in quadrature with the systematic uncertainties presented here.
\begin{table*}[tbp]
\centering
\caption{The uncertainties associated with sidelobe flux calibration. The ``Holography'' column lists the uncertainties that arise from the holography beam model described in Section \ref{sec:holo}. The $M_{\mathrm{FFT}}$ column lists uncertainties arising from the FFT-formed beam model described in Section \ref{sec:MFFT}. The ``Intensity Steady Source'' column lists the uncertainties due to the calibration constants, described in Section \ref{sec:intensity}. We use inflated errors when the hour angle of the burst exceeds $20^\circ$, since we cannot test the holography beam model in these cases.}
\begin{tabular}{lcccc}
\hline\hline
                                                          & Holography & $M_{\mathrm{FFT}}$ & Intensity Steady Source & Total Error \\ \hline\hline
\multicolumn{1}{l|}{Baseband}                             & 0.2     & --               & --                         & 0.2         \\
\multicolumn{1}{l|}{Intensity}                            & 0.2     & 0.11               & 0.07                       & 0.24        \\
\multicolumn{1}{l|}{Baseband (\textgreater{}20$^\circ$)}  & 0.4   &--                 & --                         & 0.4         \\
\multicolumn{1}{l|}{Intensity (\textgreater{}20$^\circ$)} & 0.4   &0.11                 & 0.07                       & 0.42        \\ \hline\hline
\end{tabular}
\label{tab:total_error}

\end{table*}

\section{Conclusions and Future Work}
\label{sec:conclusions}
We have outlined a method for calibrating the flux density of bursts detected in the CHIME sidelobes. In this study, we present flux calibration for both beamformed intensity data and raw, channelized baseband data. We use holography measurements between the CHIME feeds and the Galt Telescope to characterize the CHIME primary beam and the sidelobe attenuation. We isolate and validate each variable in the flux density calibration procedure, including the holography primary beam model, the FFT-formed beam model, and the calibration constants. Furthermore, we estimate the uncertainties for each of these variables in the sidelobes. This work will provide reliable flux density calibrations for FRBs, pulsars, and other astrophysical transients detected in the CHIME sidelobes. One of the main limitations of this work is the small number of holography datasets. Future work will include using holography tracks of the Sun throughout the year to provide a continuous primary-beam model over the declination range covered by the Sun ($\pm23.4^\circ$), filling in between the sparse holography sources.

\begin{acknowledgments}
\allacks
\end{acknowledgments}


\clearpage
%
\facilities{CHIME, DRAO:26m}

\software{astropy \citep{2013A&A...558A..33A,2018AJ....156..123A,2022ApJ...935..167A}}


\bibliography{main}{}
\bibliographystyle{aasjournalv7}



\end{document}